\documentclass{llncs}

\usepackage[utf8]{inputenc}
\usepackage[T1]{fontenc}
\usepackage{mathpartir}
\usepackage{wasysym}
\usepackage{listings}
\usepackage{alltt}

\title{Liveness-Driven Random Program Generation}
\author{Gergö Barany}
\institute{Inria Paris, France \\
           \email{gergo.barany@inria.fr}}

\newcommand\ldrgen{\texttt{ldrgen}}
\newcommand{\liveout}[1]{\ensuremath{#1^{\circ}}}
\newcommand{\livein}[1]{\ensuremath{#1^{\bullet}}}
\newcommand{\triple}[3]{\ensuremath{
    \left\langle#1\right\rangle \ #2 \ \left\langle#3\right\rangle
}}
\def\union{\cup}

\begin{document}
\maketitle
\thispagestyle{plain}

\begin{abstract}
Randomly generated programs are popular for testing compilers and program
analysis tools, with hundreds of bugs in real-world~C compilers found by
random testing. However, existing random program generators may generate
large amounts of dead code (computations whose result is never used). This
leaves relatively little code to exercise a target compiler's more complex
optimizations.

To address this shortcoming, we introduce liveness-driven random program
generation. In this approach the random program is constructed bottom-up,
guided by a simultaneous structural data-flow analysis to ensure that the
generator never generates dead code.

The algorithm is implemented as a plugin for the Frama-C framework. We
evaluate it in comparison to Csmith, the standard random C program
generator. Our tool generates programs that compile to more machine code
with a more complex instruction mix.
\end{abstract}


\section{Motivation}

Optimizing compilers for real-world programming languages are complex pieces
of software. Compiler bugs may manifest in several ways: As compiler
crashes, missed optimizations, or as silent miscompilations. The third
category is especially serious as it may introduce bugs in correct programs.
Such wrong-code bugs may invalidate all correctness guarantees provided by
source-level verification of safety-critical (and other) software systems.

Two main avenues of work address these problems: compiler verification and
compiler testing. Compiler verification has seen much
research~\cite{dave-2003}, with Comp\-Cert as a prominent
example~\cite{leroy-2009}. However, such compilers have not entered the
mainstream yet: Compiler verification is difficult and time-consuming, and
verified compilers therefore perform fewer optimizations and target fewer
CPU architectures than others.

A different approach is to test compilers in a way that instills confidence.
Standard compiler test suites exist for exercising C compilers, in
particular for testing their conformance to various details of the
standard~\cite{perennial,plumhall}.
In addition, randomized differential testing has gained prominence in recent
years. Compiling many random programs with various compilers and comparing the
behaviors
of the generated binaries can uncover input programs that cause compiler
crashes or miscompilations. The best-known example of this approach is the
work of Yang et al.\ on Csmith~\cite{csmith-2011}, a generator of random C
programs. Csmith generates programs that are fully self-contained (including
all their inputs in initialized global variables) and conform to the C
standard by construction. If two compilers produce code that behaves
differently for a Csmith-generated program, one of the compilers must
contain a miscompilation bug. Testing of C compilers with Csmith has
uncovered hundreds of bugs in total, including crashes and miscompilations
in every compiler under test. This included bugs in (unverified parts of)
the CompCert verified C compiler~\cite{csmith-2011}.

This article describes a random generator of C programs developed for a
project on finding missed optimizations in C compilers. Inspired by the
successes of Csmith, in this project we generate random C programs, compile
them using various compilers, then use custom tooling to search for possible
optimizations in the resulting binaries. (The details will be described in a
separate paper.) 

Starting with Csmith as our program generator, we found early on that it was
not an optimal fit for our intended use case: Despite generating
realistic-looking programs with complex arithmetic expressions, accesses to
global and local variables including through pointers, structures, and
arrays, as well as nested loops and branches, it produces large amounts
of~\emph{dead code} whose results are never used. (See our experiments in
Section~\ref{sec:evaluation}.) Dead code elimination, a
standard part of every optimizing compiler, can thus remove large parts of
the code generated by Csmith, leaving very little relevant code for the
remaining more interesting optimizations. Csmith often
generate functions of several hundred lines of code that are compiled to ten
machine instructions, representing only a small fraction of the computations
present on the source code level.

In this paper we address this problem with our new~\emph{liveness-driven}
random generator~\ldrgen. Our tool uses liveness analysis during program
generation to avoid generating dead code. In the following sections we
describe liveness-driven program generation; the implementation of our tool
as a Frama-C plugin; and its experimental evaluation, showing that~\ldrgen\
generates programs that compile to a larger amount of code and a more
complex instruction mix than programs generated by Csmith.

\section{Fully live programs}

In this section we briefly recall the basics of liveness analysis and then
present our novel formulation as a set of structural inference rules.

\subsection{Principles of liveness analysis}

A variable is called~\emph{live} at a program point if the value it holds at
that point may be read in the future, without an intervening redefinition;
otherwise, it is called~\emph{dead}.
For example, in a code snippet like \verb|x = a + b;| \verb|x = 0;|
\verb|return x;|, the variable~\verb|x| is dead after the first assignment
but live after the second one (because it is used in the \verb|return|
statement). We can extend the notion of liveness from variables to the
assignment statements defining them: An assignment \(v\)\verb| = |\(e\) is
live iff the variable~\(v\) is live just after this statement. The
first assignment to~\verb|x| above is dead, the second one is live.
(Unfortunately, some authors
use the term \emph{dead code} to refer to \emph{unreachable} code, as in
\texttt{if (false) x = y}. These concepts are not the same; our use of the
terms \emph{live} and \emph{dead} does not refer to reachability.)

Dead assignments without other side effects are useless and can be removed
from the program. Even mildly optimizing compilers implement a dead code
elimination pass that would completely remove the addition from the first
program fragment above. Our goal is to generate only live code, i.\,e., only
code that does~\emph{not} contain any such opportunities for dead code
elimination.

Liveness analysis is one of the classical data-flow
analyses~\cite{nielson.etal-1999}. It is traditionally performed as backward
fixed-point iteration over a program's control-flow graph. A statement \(S\)
in the control-flow graph has a \emph{live-in} set~\(\livein{S}\) and a
\emph{live-out} set~\(\liveout{S}\) which capture the sets of live variables
before and after execution of~\(S\). Every statement~\(S\) also has a
\emph{transfer function}~\(f_S\) relating these sets. The important case are
assignments of the form~\(v = e\) for some variable~\(v\) and
expression~\(e\). We have the following transfer function for such
assignments, where~\(FV(e)\) denotes the set of all variables occurring
in~\(e\):
\[
\livein{S} = f_{v = e}(\liveout{S})
           = (\liveout{S} \setminus \{v\}) \union FV(e)
\]

The function is said to \emph{kill} the set~\(\{v\}\) and \emph{generate}
the set~\(FV(e)\). Unless~\(v\) occurs in~\(e\), it will be dead before this
assignment, since the value it holds at that point will be overwritten by
the assignment.

Evaluation of expressions without overwriting a variable,
as in a branch condition, only generates new live variables. Individual
transfer functions are connected in a system of equations such that each
statement's live-out set is the union of the live-in sets of all of its
successors:~\(\liveout{S} = \bigcup_{S_i \in \mathit{succ}(S)}
\livein{S_i}\)

If the program contains loops, their live-out sets will depend recursively
on their own live-in sets. The desired solution is the~\emph{least fixed
point} of the system of equations, which can be found efficiently by
propagating data-flow information backward, i.\,e., starting at the
program's end and proceeding against the direction of control flow.

\subsection{Recognizing fully live programs}

We will call a program~\emph{fully live} if all of the assignment statements
it contains are live. This section develops an inference system
characterizing fully live programs.

Figure~\ref{fig:syntax} shows the abstract syntax of our programming
language of interest, a subset of C function bodies without declarations.
The language contains variables, constants, and all side-effect-free
arithmetic and bitwise operators of~C. Statements are assignments (the only
source of side effects), \verb|return| statements, \verb|if| statements and
general~\verb|while| loops. In contrast to C's concrete syntax, we view the
semicolon~\verb|;| as a statement separator, not a terminator. For now, the
language does not include~\verb|for| loops, nor any structures, arrays, or
pointers. All variables are considered local.

\begin{figure}
\[
\begin{array}{rll}
v ::= &
    \mathtt{a} \mid \mathtt{b} \mid \dots \mid \mathtt{x} \mid \mathtt{y}
    \mid \dots &
    \mbox{variables} \\
n ::= & \mathtt{0} \mid \mathtt{1} \mid \dots & \mbox{constant literals} \\
e ::= &
    v \mid
    n \mid
    e \mbox{\texttt{ + }} e \mid e \mbox{\texttt{ - }} e \mid \dots
    \mid e \mbox{\texttt{ <{}< }} e \mid \dots \mid \mbox{\texttt{-}} e \mid
    \mbox{\texttt{!}} e
    \mid \dots \qquad &
    \mbox{expressions} \\
S ::=
    & v \mbox{\texttt{\ =\ }} e & \mbox{assignment statement} \\
\mid & \mathtt{return\ } v & \mbox{return statement} \\
\mid & \mbox{\texttt{\{\}}} & \mbox{empty block} \\ 
\mid & S \mathtt{;\ } S & \mbox{sequencing} \\
\mid & \mathtt{if\ (}e\mathtt{)\ } S \mathtt{\ else\ } S & \mbox{conditional branch} \\
\mid & \mathtt{while\ (}e\mathtt{)\ } S & \mbox{loop}
\end{array}
\]
\caption{Abstract syntax of a C-like programming language}
\label{fig:syntax}
\end{figure}

Figure~\ref{fig:rules} shows a system of inference rules that characterize
fully live programs. In these rules we use a notation similar to Hoare
triples. A \emph{liveness triple} \[\triple{\livein{L}}{S}{\liveout{L}}\]
means that the variables in the set~\(\livein{L}\) are live immediately
before the statement~\(S\) (\emph{live in}), and the variables
in~\(\liveout{L}\) are live immediately after~\(S\) (\emph{live out}). A
program~\(S\) is fully live iff there is a set of variables~\(\livein{L}\)
such that the liveness triple~\(\triple{\livein{L}}{S}{\emptyset}\) is
derivable in the system.

\begin{figure}
\[
\inferrule*[left=Return]{
}{
    \triple{\{v\}}{\mathtt{return}\ v}{\emptyset}
}
\qquad
\inferrule*[left=Skip]{
    L \neq \emptyset
}{
    \triple{L}{\mathtt{\{\}}}{L}
}
\]

\[
\inferrule*[left=Assign]{
    v \in \liveout{L} \\
    \livein{L} = (\liveout{L} \setminus \{v\}) \union FV(e)
}{
    \triple{\livein{L}}{v\ \mathtt{=}\ e}{\liveout{L}}
}
\]

\[
\inferrule*[left=Sequence]{
    \triple{\livein{L_1}}{S_1}{\livein{L_2}} \\
    \triple{\livein{L_2}}{S_2}{\liveout{L_2}} \\
    \livein{L_2} \neq \emptyset
}{
    \triple{\livein{L_1}}{S_1\ \mathtt{;}\ S_2}{\liveout{L_2}}
}
\]

\[
\inferrule*[left=If]{
    \triple{\livein{L_1}}{S_1}{\liveout{L}} \\
    \triple{\livein{L_2}}{S_2}{\liveout{L}} \\
    \livein{L} = \livein{L_1} \union \livein{L_2} \union FV(c)
}{
    \triple{\livein{L}}
           {\mathtt{if\ (}c\mathtt{)}\ S_1\ \mathtt{else}\ S_2}
           {\liveout{L}}
}
\]

\[
\inferrule[While]{
    \triple{\livein{B}}{S}{\liveout{B}} \\
    \liveout{B} = \liveout{L} \union \livein{B} \union FV(c)
        \mbox{ (minimal)} \\
    \livein{L} = \livein{B} \union \liveout{L} \\
    \liveout{L} \neq \emptyset
}{
    \triple{\livein{L}}
           {\mathtt{while\ (}c\mathtt{)}\ S}
           {\liveout{L}}
}
\]
\caption{System of inference rules for fully live programs}
\label{fig:rules}
\end{figure}

Intuitively, the system of inference rules encodes two things. First, the
rules are an alternative presentation of the transfer functions of liveness
analysis. A triple~\(\triple{\livein{L}}{S}{\liveout{L}}\) that appears in a
valid derivation corresponds to a data-flow equation~\(\livein{L} =
f_S(\liveout{L})\) where~\(f_S\) is the transfer function for the
statement~\(S\). For example, the transfer function~\(f_{v = e}\) for an
assignment~\(v\)\verb| = |\(e\) is just~\(f_{v = e}(\liveout{L}) =
(\liveout{L} \setminus \{v\}) \union FV(e)\), as in the side condition of
the~\textsc{Assign} rule. Similarly, the~\textsc{Sequence} rule encodes the
composition of transfer functions, and the~\textsc{If} rule encodes the
split and join of data-flow information along different program paths.

Second, the other side conditions add a system of constraints to ensure full
liveness: Any assignment statement appearing in a fully live program~\(S\)
(i.\,e., for which a derivation of~\(\triple{\livein{L}}{S}{\emptyset}\) for
some~\(\livein{L}\) exists) is itself live. This follows directly from
the~\textsc{Assign} rule's side condition~\(v \in \liveout{L}\).
For example, a triple of the form
\[
\triple{\livein{L}}{\mbox{\texttt{x = a; x = b}}}{\liveout{L}}
\]
can never be derived in the system because the first of the two assignments
is dead. The~\textsc{Sequence} rule says that to derive this triple, there
must be some intermediate set~\(L'\) of variables such that \(\mathtt{x} \in
L'\) due to \textsc{Assign} on~\verb|x = a| while at the same time
\(L' = (L \setminus \{\mathtt{x}\}) \union \{\mathtt{b}\}\) due
to~\textsc{Assign} on~\verb|x = b|. This is a contradiction, so the
derivation attempt must fail.

While the other rules are straight-forward, the \textsc{While} rule deserves
some discussion. Unlike the two branches of the~\verb|if| statement, the
whole loop's live-out set~\(\liveout{L}\) is not identical to the loop
body's live-out set~\(\liveout{B}\): Typically there are loop-carried
dependences, i.\,e., cases where a variable is set on one iteration of the
loop and its value is read on a later iteration. Such variables are live out
of (and live into) the loop body, but if they are no longer used once the
loop has terminated, they are not live out of the loop. When performing a
derivation in the system, we must guess or calculate the set of these
additional variables.

Let~\(f_S\) denote the liveness transfer
function corresponding to the loop body statement~\(S\). Then from the
liveness triple~\(\triple{\livein{B}}{S}{\liveout{B}}\) we have~\(\livein{B}
= f_S(\liveout{B})\), and the equation~\(\liveout{B} = \liveout{L} \union
\livein{B} \union FV(c)\) means that~\(\liveout{B}\) is a fixed point of the
function~\(\lambda\,B.\;(\liveout{L} \union f_S(B) \union FV(c))\). The
minimality side condition additionally specifies that we are interested in
the \emph{least} fixed point of this function. This least fixed point exists
and is unique~\cite{nielson.etal-1999}.

\begin{figure}
\[
\inferrule*{
\inferrule*{
\inferrule*{
    \inferrule*{
        \mathtt{
        \triple{\{a, b, n\}}
               {\mbox{\texttt{t = a + b}}}
               {\{b, n, t\}}} \\
        \inferrule*[vdots=1.35em, leftskip=7.5em]{
            \mathtt{
            \triple{\{b, n, t\}}
                   {\mbox{\texttt{a = b}}}
                   {\{a, n, t\}}} \\
            \inferrule*[vdots=1.35em, leftskip=10em]{
                \mathtt{
                \triple{\{a, n, t\}}
                       {\mbox{\texttt{b = t}}}
                       {\{a, b, n\}}} \\
                \inferrule*[vdots=1.35em, leftskip=10em]{
                    \mathtt{
                    \triple{\{a, b, n\}}
                           {\mbox{\texttt{n = n - 1}}}
                           {\{a, b, n\}}}
                }{ }
            }{ }
        }{ }
    }{
        \mathtt{
        \triple{\{a, b, n\}}
               {\mbox{\texttt{t = a + b;\ a = b;\ b = t;\ n = n - 1}}}
               {\{a, b, n\}}}
    }
}{
    \mathtt{
    \triple{\{a, b, n\}}
           {\mbox{\texttt{while\ (n\,>\,0)\ %
                  \{\ t\,=\,a\,+\,b;\ a\,=\,b;\ b\,=\,t;\ n\,=\,n\,-\,1\ \}}}}
           {\{a\}}}
}
}{
    \vdots
}}{
    \mathtt{
    \triple{\{n\}}
           {\mbox{\texttt{a\,=\,0;\,b\,=\,1;\,%
                    while\,(n\,>\,0)\,\{\,t\,=\,a\,+\,b;\,a\,=\,b;\,b\,=\,t;\,n\,=\,n\,-\,1\,\};\,%
                    return\,a}}}
           {\emptyset}}
}
\]
\caption{Example derivation proving full liveness.}
\label{fig:example}
\end{figure}

In Figure~\ref{fig:example} we illustrate the use of the inference system to
prove full liveness of a program taking an input variable~\texttt{n}
(assumed to be non-negative) and returning the~\verb|n|-th Fibonacci number.
We omit some details to focus on the analysis of the loop. Note that only
the return variable~\verb|a| is live after the loop. However, the live-out
set of the loop's body is~\(\{\mathtt{a, b, n}\}\). This includes the return
variable~\verb|a| and the variable~\verb|n| that is used in the loop
condition. It also includes the variable~\verb|b|, which is the element
computed by fixed point iteration: The value of~\verb|b| at the end of the
loop body will be used on the next loop iteration, if any. Conversely,
if~\verb|b| were not live at some point in the loop body, our inference
system would not allow derivation of a triple for the assignment~\verb|b = t|.
Indeed, all assignments in the loop body satisfy the condition that they
define variables that are live after the assignment. That is, this program
is~\emph{fully live} by our definition.

\subsection{Limitations of the system}

Note that fully live programs may still contain opportunities for simple
optimizations that remove code that does not have interesting effects. For
example, programs accepted by the inference system above may contain
fragments like
    \verb|if (0) { ... } else { ... }|
where one of the branches of the~\verb|if| statement is unreachable and thus
irrelevant; or assignments like~\verb|y = ...; x = y - y;| where the
computation for the value of~\verb|y| is irrelevant for~\verb|x|'s final
value of~0. Our inference rules do not consider the semantics of the code in
enough detail to exclude such cases.

Our claims with regards to full liveness are relative to a purely syntactic
notion of liveness that does not consider such semantic issues. In
particular, we cannot guarantee that the liveness analysis embedded in
these rules is equivalent to liveness analysis as performed by any given
compiler. Any other analyses or transformations performed by the compiler
before liveness analysis may influence the results, typically making the
compiler's results more precise than ours.

As our experimental results in Section~\ref{sec:evaluation} show, our
generator performs well nonetheless, so we can leave refinements of the
system for future work.

\subsection{Generating fully live programs}

The inference rules can be
translated almost directly into an executable random (or exhaustive)
generator of fully live programs. Like traditional liveness analysis,
generation proceeds backwards, i.\,e., in the direction opposite control
flow.

The side conditions of the inference rules ensure that a fully live program
always ends in a return statement, as no other statement may have an empty
live-out set. The generator can thus start by picking a random program
variable~\(v\) and generating a statement~\(\mathtt{return}\ v\) with
live-in set~\(L = \{v\}\). It then iteratively prepends random
statements~\(S\) to the current program fragment and updates the live-in set
according to~\(f_S(L)\). The possibilities for the generation of~\(S\) are
guided by~\(L\). In particular, if the generator decides to generate an
assignment statement, the target variable~\(v\) must be an element of~\(L\)
at that point. Conversely, if~\(L\) ever becomes empty, generation of the
current block of code must stop at that point: Any code preceding that point
would be dead. Figure~\ref{fig:generator} shows pseudocode of such a
generator in an OCaml-like functional language.

\begin{figure}
\lstset{emph={v,L,S,e,code,c,t,f,V,B}}
\begin{lstlisting}
let random_statements L code =
    if L $\neq \emptyset$ then
        let (S, $L'$) = random_statement L;
        random_statements $L'$ (S :: code)
    else
        (code, L)

let random_statement L =
    let statement_generator = random_select [assignment; branch; loop];
    statement_generator L

let assignment L =
    let v = random_select L;
    let e = random_expression ();
    ("$v$ = $e$", $(L \setminus \{v\}) \union FV(e)$)

let branch L =
    let (t, $L_1$) = random_statements L [];
    let (f, $L_2$) = random_statements L [];
    let c = random_expression ();
    ("if ($c$) $t$ else $f$", $L_1 \union L_2 \union FV(c)$)

let loop L =
    (* See main text for explanation. *)
    let c = random_expression ();
    let $B'$ = random_variable_set ();
    let (code, L') = random_statements $(L \union B' \union FV(c))$ [];
    let $V = \{ b \in B' \mid b \notin L' \mbox{ or \(b\) not used in \(S\)}\}$;
    if $V \neq \emptyset$ then
        let e = random_expression_on_variables V;
        let v = random_select $L'$;
        let code' = "$v$ = $e$" :: code;
        ("while ($c$) $code'$", $(L' \setminus \{v\}) \union V \union L$)
    else
        ("while ($c$) $code$", $L' \union L$)

(* Start generation with the terminating return statement. *)
let v = random_variable ()
let (code, L) = random_statements {v} ["return $v$"]
\end{lstlisting}
\caption{Pseudocode of a liveness-driven random program generator.}
\label{fig:generator}
\end{figure}

Every statement generation function takes a live variable set~\(L\) and
returns a pair of a newly generated statement and an updated live variable
set according to the statement's transfer function.
We iterate statement generation and collect a list of statements
forming a block. In this presentation we omit helper functions such as the
ones for generating random variables and expressions.

As before, the handling of loops merits more discussion. Just as the
inference system needs a minimal set~\(\livein{B}\) containing new variables
that are live into and out of the loop body, the random generator must
synthesize such a set of variables. But here the problem is more difficult:
During inference, we can start with an initial live-out set and derive the
eventual live-in set by fixed-point iteration. During code generation this
is not possible since we cannot analyze the loop body before we have
constructed it. Instead, we first generate a random set of newly live
variables and let this choice guide generation of the loop's code.

This solution relies on the following observation: The new variables we are
interested in are ones that are defined before the loop, may be defined on
some loop iteration, and then used on some later iteration. In the example
of Figure~\ref{fig:example}, this is the case for variable~\verb|b|, which
stores the next Fibonacci number for assignment to~\verb|a| on the
subsequent iteration. Using the names in the~\textsc{While} rule of
Figure~\ref{fig:rules}, the set of these `new' variables is~\(B' =
\liveout{B} \setminus (\liveout{L} \union FV(c))\). It follows that~\(B'
\subseteq \livein{B}\), i.\,e., every~\(b \in B'\) is live into the loop.

To generate a fully live loop, we choose a random set~\(B'\) of new
variables and generate a loop body in a way that ensures that~\(\liveout{B}
= \liveout{L} \union B' \union FV(c)\) is a least fixed point of the loop.
For this we add~\(B'\) to the live variable set~\(L\)
before generating a loop body block (along with the live variables~\(FV(c)\)
generated by the loop condition). The generated loop body may or may not
define or use variables in~\(B'\), and thus these variables may or may not
be live into the generated block. However, we must force them to be used in
the loop body and be live into the body: If there were some~\(b\) in the
generated~\(B'\) that is not live into the loop body, we would violate the
condition~\(B' \subseteq \livein{B}\) established above. On the other hand,
if~\(b\) were live into the loop body but did not have a use anywhere in the
body, then~\(\livein{B}\) would not be minimal and hence~\(\liveout{B}\)
would not be a least fixed point of the constraint system.

To ensure a correct, minimal solution, we therefore find the set~\(V\) of
all~\(b \in B'\) that are not
in~\(L'\) or that are in~\(L'\) but have no use in the generated loop body.
We pick a random live variable \(v \in L'\) and prepend an
assignment~\(v\)\verb| = |\(e\) to the generated loop body, where~\(e\) is
an expression containing all the variables in~\(V\). This final loop body
ensures that all variables in~\(B'\) are live into it and used in it, hence
ensuring that~\(\liveout{B} = \liveout{L} \union B' \union FV(c)\) is a
least fixed point of the loop's liveness constraint system.

A small detail not illustrated in the pseudocode is the case when~\(L'\) is
empty at the beginning of the generated loop body. This can only be the case
if the first statement in the body is an assignment of a constant expression
(i.\,e., not using any variables) to some variable~\(v\), since such
assignments are the only statements that can remove variables from the live
variable set without adding any new ones. In this case, we replace this
assignment's right-hand side with the expression~\(e\) generated as above.

\section{Implementation}

We have implemented the random program generation algorithm sketched above
as a plugin for Frama-C~\cite{frama-c}. Frama-C is a general, extensible
framework for source-level analysis and transformation of C programs. It is
written in OCaml and can be extended with plugins written in that language.
For this work, we do not need any advanced Frama-C features but benefit from
its AST (abstract syntax tree) type definitions, utilities for managing
variables and constructing AST fragments, and its pretty-printer for
outputting the generated AST as C source code. As these general parts are
provided by Frama-C, \ldrgen\ itself can be quite small: It consists of only
about~600 lines of generator code, plus some utilities and configuration.

\ldrgen\ is free software, available at
\texttt{\url{https://github.com/gergo-/ldrgen}}.

\subsection{Random generation}

The core of the generator has the same structure as the pseudocode in
Figure~\ref{fig:generator}. After generating an empty function definition
and a return statement, it fills in the function's body by generating a
fully live sequence of statements as in the pseudocode. Statements are
represented by AST fragments; we never need to worry about generating actual
C syntax. The current version of \ldrgen\ always generates a single
function.

Random expressions are generated by choosing an operator among the
arithmetic operators available in C and recursively generating the
appropriate number of operand expressions. At this point, C's type system
becomes relevant; if needed, we insert type casts to ensure all operands of
an operator have the same type. Type casts are also needed in some other
cases: Bitwise operators and the modulo operator cannot be applied to
floating-point numbers in C, so we insert conversions to integer types in
such cases.

Many C operators may invoke undefined behavior when applied to inappropriate
values. Two examples are division by zero and signed integer arithmetic
overflow. Unlike Csmith in its default mode, \ldrgen\ does not try to guard
against such undefined operations, except for two cases that compilers have
repeatedly warned us about: We clamp the right-hand-side operands of
bit-shift operations to the bit size of the expression on the
left-hand-side, and we always generate division and modulo operations of the
form~\(e_1 \mathtt{/} \mathtt{(}e_2 \mathtt{+} c\mathtt{)}\) for some
constant~\(c\) instead of just~\(e_1 \mathtt{/} e_2\). The idea behind this
is that~\(e_2 \mathtt{+} c\) is less likely to evaluate to zero than a
random expression in general. This approach is primitive, but we have found
it to work well in practice.

Leaves of expressions are constants or variables. For constant literals we
simply generate a random number. For a variable use we either use a
previously used variable or generate a new one. Variables generated in
this way may be local variables or function parameters. Both can be used in
expressions, but we only generate assignments to locals, not to parameters.

Some of the generator's choices are weighted by manually chosen parameters
to ensure generation of somewhat more realistic-looking programs. For
example, we prefer generation of basic arithmetic operations to bitwise
operators. We also ensure that loop and branch conditions are not constant
expressions, i.\,e., that they contain at least one variable. In order to
avoid trivial non-termination issues, we also ensure that every loop body's
final statement is an assignment to some variable that occurs in the loop
condition. If there were no modification of any of these variables at all, a
loop once entered could never terminate. Even so, termination is not at all
guaranteed.

Bottom-up generation of the function's body may stop if there are no more
live variables, or if a user defined limit is reached. In this
latter case, there may remain live local variables at the start of the
function's body. Their liveness means that they may be used without being
assigned to, so we must ensure that they are initialized. We therefore
finalize the function definition by initializing all such live-in variables
to constants or to the values of function parameters.

\subsection{Configuration}

The generator's behavior may be tuned using command-line arguments. These
may specify features of the sub-language of C that is used. For example, the
user may request the generation of code that only uses integer types, or
only floating-point types. They may also specify that no bitwise operations
or no divisions should be generated, and whether loops may be generated.
Other flags specify structural properties: The maximal number of statements
per block, and the maximal nesting depths of statements and expressions.

\ldrgen's random generation uses OCaml's standard pseudorandom number
generator, which can be seeded with a random seed or with a seed value
specified as a command line argument. Invoking a given version of \ldrgen\
with a fixed set of arguments and a fixed seed thus always gives the same
reproducible result.

\subsection{Extensions to the basic model}

We describe two extensions to the core language of Figure~\ref{fig:syntax}
that are already implemented in \ldrgen: very limited uses of pointers
and~\verb|for| loops over arrays.

First, in addition to the arithmetic types used so far, we can generate
function parameters of type~\verb|T *| (pointer to~\verb|T|) for some
arithmetic type~\verb|T|. A parameter~\verb|p| of such type can be used in
generated code as~\verb|*p|. We currently do not generate assignments to
such dereferenced pointers, nor any pointer arithmetic.

Second, we want to generate arrays and restricted forms of loops over them
in order to exercise loop optimizations such as unrolling or vectorization.
For this we generate pointer arguments~\verb|T *arr| which are only
used in~\verb|for| loops of the following form:

\begin{quote}
\begin{alltt}
v = \(\dots\);
for (unsigned int i = 0; i < N; i++) \{
    v = v \(\circ\) \(f(\mbox{\texttt{arr[i]}})\);
\}
\end{alltt}
\end{quote}

Here, \verb|N| is a global variable considered to hold the array's
size,~\(f(\mbox{\texttt{arr[i]}})\) is a random expression
involving~\verb|arr[i]|, and~\(\circ\) is a randomly chosen binary
arithmetic operator. This loop pattern implements a map-reduce operation,
mapping some function~\(f\) over the array and reducing (folding) the result
with~\(\circ\). It is currently the only kind of~\verb|for| loop we
implement, but this would be easy to generalize.
Similar forms of loops are already generated by Csmith, but there their
results are virtually never used. In~\ldrgen, we choose a loop result
variable~\verb|v| that is live after the loop to ensure that it is used.

\subsection{Future extensions}

In the future, we are planning to extend \ldrgen\ to generate structure
types and allow the use of their members.

In the longer term, \ldrgen\ will also be extended to support programs
consisting of several random functions which may call each other. We are not
planning to support non-structured control flow using~\verb|goto|. The more
structured~\verb|break| and~\verb|continue| statements might eventually be
supported, but this is not a priority as they complicate the structural
liveness analysis.

\section{Evaluation}
\label{sec:evaluation}

The design goal of~\ldrgen\ was to have a random program generator that
exposes as much interesting code as possible to all passes of the compiler
under test; recall that we found Csmith-generated code to contain much dead
code which is never seen by many parts of the compiler because it can be
optimized away early on. We will have achieved our goal if, for comparable
amounts of generated C code, \ldrgen's output results in more, and ideally
more varied, assembly code than Csmith's output. We therefore compare the
two generators along these lines. We do not claim superiority to Csmith in
any other regard, especially not concerning its power to find subtle
miscompilation bugs.

Csmith is designed to run complete, self-contained applications consisting
of several functions, driven by a~\verb|main| function. In contrast,
\ldrgen\ only generates individual functions without a driver. However,
Csmith's many configuration options allow us to ask it to generate files
consisting only a single function without~\verb|main|.\footnote{The concrete
flags we used were \texttt{--nomain --float --max-funcs 1 --no-safe-math
--max-block-size 8 --concise}.}

Table~\ref{tab:results} presents our experimental results for~1000 programs
each generated by Csmith and \ldrgen. We investigate three characteristics
of the generated programs: lines of C code, number of instructions in the
generated code, and number of unique opcodes in the generated code. In all
cases, the C code was compiled to x86-64 machine code using GCC 5.4.0 with
optimization setting~\verb|-O3|. For each characteristic, the table shows
the total over the~1000 files as well as the minimum, median, and maximum
values. (In cases where the median is not unique, we chose the arithmetic
mean of the two closest values.)

\begin{table}
\caption{Comparison of code generated by Csmith and \ldrgen\ in 1000 runs
each.}
\label{tab:results}

\centering
\begin{tabular}{l@{\ } | @{\ }l @{\quad} r @{\quad} r @{\quad} r @{\quad} r}

& generator & min & median & max & total \\
\hline
%
%
%
lines of code & Csmith
    & 25 & 368.5 & 2953 & 459021 \\
& \ldrgen
    & 12 & 411.5 & 1003 & 389939
    \\[1ex]
instructions & Csmith
    & 1 &  15.0 & 1006 &   45606 \\
& \ldrgen
    & 1 & 952.5 & 4420 & 1063503
    \\[1ex]
unique opcodes & Csmith
    & 1 &  8 &  74 & 146 \\
& \ldrgen
    & 1 & 95 & 124 & 204
    \\
%
%
\end{tabular}
\end{table}



Our command line flags for Csmith were chosen in order to generate
comparable numbers of lines of C code to \ldrgen. In fact
Table~\ref{tab:results} shows that it generates somewhat more, but these
numbers are difficult to compare precisely because Csmith-generated code
tends to contain many initializers for global variables; \ldrgen\ does not
generate any global variables at all. We believe that the settings we chose
allow a fair comparison of the generators.

Next we compare the number of instructions (executable code only, excluding
static data, assembler directives etc.) emitted by the compiler for the
generated source files. \ldrgen\ was designed to increase this number
compared to Csmith, and the table shows that we have succeeded: While on
average Csmith's output compiles to a single machine instruction per ten
lines of code, \ldrgen's output has almost three instructions per single
line of source code. Overall, \ldrgen-generated programs compile to about 20
times as much machine code as Csmith-generated programs of comparable size.
We can also see that the distribution for Csmith is highly skewed: The
median shows that at least half of the functions generated by Csmith compile
to~15 instructions or fewer. This also confirms our initial, more informal
observation that Csmith-generated code tends to contain large amounts of
dead code. \ldrgen\ manages to generate code with a less skewed
distribution, and in particular with generally higher numbers of emitted
instructions.

On a side note, we remark that both Csmith and \ldrgen\ sometimes generate
functions that compile to a single machine instruction. Inspection showed
that this happens in cases where the compiler recognizes that a function
ends up in an infinite loop without externally visible side effects. Such
functions are then compiled into a single unconditional jump instruction
looping back to itself. Many other functions compile to two instructions,
typically some simple operation on a function argument or a constant
followed by a return. It would be difficult to completely avoid generating
infinite loops, but comparatively easy (at least within \ldrgen) to avoid
generating functions that return after a single operation. For both Csmith
and \ldrgen, about~10\,\% of all cases fall into one of these trivial
categories (with Csmith producing fewer infinite loops).

We analyze the coverage of the instruction set in the generated
code by looking at the number of different opcodes generated. Here, too, we
see that individual functions generated by \ldrgen\ have a more varied
instruction mix than functions generated by Csmith: Even the median for
\ldrgen\ is higher than the maximum for Csmith. Totaling over all the
machine code in~1000 functions, we see that Csmith-generated code compiles
to a mix of~146 different opcodes, while \ldrgen-generated code contains~204
different opcodes, an increase in instruction set coverage of~40\,\%.
Inspection of the sets of opcodes shows that this difference is almost
entirely due to various vector (SIMD) arithmetic instructions generated for
\ldrgen's code. Compiling to such instructions was the goal of
adding~\verb|for| loops over arrays to \ldrgen.  Manual inspection of some
cases shows that such loops are indeed the origin of these instructions.
Disabling generation of~\verb|for| loops in~\ldrgen\ brings its total number
of unique instructions down to~147, comparable to Csmith.

One of the few opcodes emitted for Csmith-generated code but not
for~\ldrgen\ are~\verb|call| instructions to~\verb|memcpy| which are
sometimes generated by compilers for structure copies. \ldrgen\ currently
does not generate structures at all.

Finally we compare the speed of the two generators. Generating the~1000
files each analyzed above took 871~seconds with Csmith and~124 seconds with
\ldrgen\ (Csmith backtracks if it finds that it has generated unsafe code).
Csmith generates about~527 lines of C code per second, with \ldrgen\
generating~3140 (about~\(6\,\times\) more). With respect to final machine
code, Csmith-generated code compiles to about~52 instructions per second of
generation time, whereas \ldrgen\ produces~8563 (about~\(160\,\times\)).



\section{Related work}

The best-known random program generator is Csmith~\cite{csmith-2011}, based
on an earlier system
called~\texttt{randprog}~\cite{eide.regehr-2008,randprog-2007}. Csmith
generates complete, self-contained programs that take all their input from
initialized global variables and compute an output consisting of a hash over
the values of all global variables at the end of execution. The generator is
designed to only generate programs with well-defined semantics: Operations
that may be undefined in C, such as overflowing signed integer arithmetic,
are guarded by conditionals that exclude undefined cases (these guards can
be disabled, and we disabled them for the experiments reported above). Like
\ldrgen, Csmith performs data-flow analysis during generation, although the
details differ due to the differing design goals. Csmith's forward analysis
computes points-to facts and uses them for safety checks. If the checks
fail, Csmith backtracks, deleting code it generated until a safe state is
reached again. In contrast, \ldrgen's data-flow analysis only deals with
liveness, and \ldrgen\ never backtracks: Full liveness of variables in loops
is ensured by construction. Csmith generates a larger subset of C than
current or currently planned versions of \ldrgen, including unstructured
control flow and less restricted use of pointers.

Csmith has been used to find hundreds of bugs in C compilers when compiling
the programs it generates~\cite{csmith-2011}. It has also been used as the
basis of mutation-based systems, where Csmith's output was modified using
other tools to provoke compiler bugs~\cite{le.etal-2014}. The CLsmith tool
derived from Csmith has been used to find many bugs in OpenCL
compilers~\cite{lidbury.etal-2015}. Another notable generator is jsfunfuzz
for JavaScript, which has found thousands of bugs in JavaScript
engines~\cite{jsfunfuzz}.


The JTT program generator~\cite{zhao.etal-2009} is aimed directly at testing
compiler optimizations. It uses a model-based approach, where generation is
guided using test scripts. These scripts contain code templates and temporal
logic specifications of the optimizations to be tested. For example, the
authors specify opportunities for dead code elimination as cases where a
variable is assigned, then assigned again before being used. The test script
contains a temporal logic formula expressing this pattern and the test
condition that the compiler should eliminate the first assignment. Using
this script, JTT generates test programs containing this pattern. JTT was
used successfully to find bugs and increase the test suite's statement
coverage for an industrial C compiler.

Other work specifically aimed at testing and comparing program verification
tools generates code from randomly generated LTL
formulae~\cite{steffen.etal-2014}. The generated code is guaranteed to
satisfy the specified temporal properties.

Our formulation of liveness analysis as set of structural inference rules is
inspired by formulations of interval-based data-flow analysis where
reducible programs are decomposed into components called \emph{intervals},
and analysis data is efficiently propagated among the
intervals~\cite{allen-1970,cocke-1970,graham.wegman-1976}.

\section{Conclusions}

We presented~\ldrgen, a new generator of random C programs designed for
testing C compilers. In contrast to Csmith, the dominant player in this
field, \ldrgen\ is driven by liveness analysis to avoid generating dead
code. We designed an inference system to capture our liveness analysis and
implemented its rules as an excutable program generation system.

\ldrgen\ is implemented as a plugin for the Frama-C framework. Our
evaluation of \ldrgen\ in comparison to Csmith shows that we have achieved
our goal of generating C code that compiles to larger amounts of machine
code with a more interesting instruction mix, including many SIMD
instructions. We are actively using~\ldrgen\ in a project on finding missed
optimizations in compilers. Because it is able to exercise loop
optimizations not usually addressed by Csmith, it may also be useful for
finding correctness bugs in these optimizations.

\section*{Acknowledgments}

The author would like to thank the anonymous reviewers and John Regehr for
insightful comments on earlier versions of this paper.
This research was partially supported by ITEA~3 project no.~14014, ASSUME.

\bibliography{lopstr2017}
\bibliographystyle{splncs03}

\end{document}